\newcommand{\dent}{\hat \delta}
\documentclass{PoS}

\title{Holographic Flavored Quark-Gluon Plasmas}

\ShortTitle{Holographic Flavored QGPs}

\author{Francesco Bigazzi\\ Dipartimento di Fisica e Astronomia, Universit\'a di Firenze and I.N.F.N. - sezione di Firenze;
Via G. Sansone 1, I-50019 Sesto Fiorentino (Firenze), Italy.\\
        E-mail: \email{bigazzi@fi.infn.it}}

\author{\speaker{Aldo L. Cotrone}\thanks{Preprint number: DFTT 2/2011. F. B. and A. L. C. would like to thank the Italian students, parents, teachers and scientists for their activity in support of public education and research.}\\
        Dipartimento di Fisica Teorica, Universit\`a di Torino and I.N.F.N. - sezione di Torino
Via P. Giuria 1, I-10125 Torino, Italy.\\
        E-mail: \email{cotrone@to.infn.it}}

\author{Daniel Mayerson\\
        Institute for theoretical physics, K.U. Leuven;
Celestijnenlaan 200D, B-3001 Leuven,
Belgium.\\
        E-mail: \email{drm56@cam.ac.uk}}

\author{Angel Paredes\\
        Departament de Fisica Fonamental and Institut de Ciencies del Cosmos (ICC), Universitat de
Barcelona (UB), Marti Franques 1, E-08028 Barcelona, Spain.\\
        E-mail: \email{aparedes@ffn.ub.es}}

\author{Javier Tarr\'\i o\thanks{Preprint number: ITP-UU-11/03}\\
        Institute for Theoretical Physics, Universiteit Utrecht, 3584 CE, Utrecht, The Netherlands.\\
        E-mail: \email{l.j.tarriobarreiro@uu.nl}}

\abstract{Holography provides a novel method to study the physics of Quark Gluon Plasmas, complementary to the ordinary field theory and lattice approaches. In this context, we analyze the informations that can be obtained for strongly coupled Plasmas containing dynamical flavors, also in the presence of a finite baryon chemical potential. In particular, we discuss the jet quenching and the hydrodynamic transport coefficients.}

\FullConference{The many faces of QCD\\
        November 2-5, 2010\\
        Gent Belgium}

\begin{document}

\section{Introduction: the string/field theory correspondence}

The AdS/CFT correspondence implements the holographic principle,
realizing the idea that a quantum field theory in $d$ dimensions is
equivalent to a theory containing gravity in at least $d+1$
dimensions. Thus, the latter is supposed to describe all the degrees
of freedom, and the corresponding interactions, of the quantum field
theory, at least in some regimes of parameters. The master example
is the conjectured equivalence between $4d$ ${\cal N}=4$ SYM with
gauge group $SU(N_c)$ and Type IIB string theory on the $AdS_5
\times S^5$ background \cite{malda}. The main interest of this
statement resides in the parameter regime of the field theory where
the gravity approximation to string theory is reliable: the planar
limit $N_c \gg 1$ at large 't Hooft coupling $\lambda \sim g_{YM}^2
N_c \gg 1$. That is, affordable computations in a classical theory
of gravity completely determine the strong coupling regime of a
quantum field theory.

The concrete realization of this correspondence requires that each ingredient in the field theory (FT) has to have a dual description in the gravity theory.
This determines a dictionary between the two theories.
Some of its main entries are the following:\footnote{We are quite cavalier with details.}
\begin{itemize}
\item For each operator ${\cal O}$ in the FT there must be a dual gravity field $\Phi$.
\item A vacuum in FT corresponds to a background gravity solution.
\item The RG scale is dual to some radial variable $r$ in the gravity background: an asymptotically Anti de Sitter manifold with a natural boundary at $r\to\infty$.
\item Deconfined phases at finite temperature and charge density are
realized by charged black holes.
\item Transport coefficients, i.e. the response to external perturbations in FT, are derived by perturbing gravity fields.
\end{itemize}
The actual way to compute correlation functions in FT from gravity is encoded in the basic formula \cite{gkp,witten}
\begin{equation}
\langle e^{-\int \Phi_0 {\cal O}}\rangle_{FT} = e^{S_{gravity}(\Phi_0)}\, ,
\end{equation}
where $\Phi_0\equiv \lim_{r \rightarrow \infty} \Phi$ is the boundary value of the gravity field $\Phi$ dual to the operator ${\cal O}$. Via the gravity equations of motions and selecting an appropriate behavior of $\Phi$ in the interior,
the value of $\Phi$ is determined everywhere by $\Phi_0.$
The left hand side of the formula above is the generating functional in FT of the correlations functions of the operator ${\cal O}$, so essentially solves the FT; the right hand side includes the gravity action on-shell on the solution of the equations of motion for $\Phi$.

As an example of this formalism, let us consider the strongly coupled ${\cal N}=4$ SYM plasma at temperature $T$. The thermodynamically favored phase, corresponding to deconfined matter transforming in the adjoint representation, is described by a black hole in $AdS$.
Let us also consider one particular operator: the $T_{xy}$ component of the energy-momentum tensor.
Its dual gravity field turns out to be the $g_{xy}$ component of the metric.
Then, the two-point function
\begin{equation}
\langle [T_{xy}(t, \vec x),T_{xy}(0, \vec 0)] \rangle
\end{equation}
can be calculated from the on-shell supergravity action for the graviton $g_{xy}$, determined by its equation of motion on the $AdS$ black hole.

At long distances, late times as compared to $1/T$, the ${\cal N}=4$ SYM plasma admits as usual a hydrodynamic description, whose basic equation is simply the energy-momentum conservation
\begin{equation}
\nabla_\mu T^{\mu\nu}=0\, .
\end{equation}
Up to second order in the derivative expansion, the hydrodynamics energy-momentum tensor reads \cite{baier,romatschke}
\begin{equation}\label{enmom}
T^{\mu\nu}=\varepsilon u^\mu u^\nu + p  \Delta^{\mu\nu} + \pi^{\mu\nu} + \Delta^{\mu\nu}\Pi   \, ,
\end{equation}
where $\varepsilon$ is the energy density, $u^\mu$ the velocity field, $p(\varepsilon)$ the pressure, $\Delta^{\mu\nu} = h^{\mu\nu}+u^\mu u^\nu$ with $h^{\mu\nu}$ the $4d$ metric and
\begin{eqnarray}
\pi^{\mu\nu} &=&- \eta \sigma^{\mu\nu} +\eta \tau_\pi \Bigl[\langle D \sigma^{\mu\nu}\rangle + \frac{\nabla \cdot u}{3}\sigma^{\mu\nu} \Bigr] + \kappa \Bigl[ R^{<\mu\nu>}-2 u_\alpha u_\beta R^{\alpha <\mu\nu> \beta} \Bigr] \nonumber \\
&& + \lambda_1 \sigma^{<\mu}_{\lambda} \sigma^{\nu>\lambda} + \lambda_2  \sigma^{<\mu}_{\lambda} \Omega^{\nu>\lambda} + \lambda_3 \Omega^{<\mu}_{\lambda} \Omega^{\nu>\lambda}  + \kappa^* 2u_\alpha u_\beta R^{\alpha <\mu\nu> \beta}  \nonumber \\
&& + \eta \tau_\pi^* \frac{\nabla \cdot u}{3}\sigma^{\mu\nu} + \lambda_4 \nabla^{<\mu} \log{s}  \nabla^{\nu >} \log{s} \, ,
\end{eqnarray}
\begin{eqnarray}
\Pi &=&- \zeta (\nabla \cdot u) + \zeta \tau_\Pi  D  (\nabla \cdot u) + \xi_1 \sigma^{\mu\nu}\sigma_{\mu\nu}+ \xi_2 (\nabla \cdot u)^2 + \xi_3 \Omega^{\mu\nu}\Omega_{\mu\nu} \nonumber \\
&&  + \xi_4 \nabla_{\mu}^{\perp} \log{s} \nabla^{\mu}_{\perp} \log{s}+ \xi_5 R + \xi_6  u^\alpha u^\beta R_{\alpha \beta}\, .
\end{eqnarray}
For the precise definition of the structures in these expressions we refer to \cite{romatschke}.
In principle, the transport coefficients of the structures above are determined by
the microscopic theory.
The most important transport coefficients are the shear and bulk viscosities $\eta, \zeta$ and the two relaxation times $\tau_\pi, \tau_\Pi$, which have possible implications for the elliptic flow measured at RHIC and LHC.

For a general strongly coupled theory, the theoretical determination of the transport coefficients is a daunting task.
In the case at hand, on the contrary, they can be extracted with a reasonable amount of work from gravity.
The shear viscosity, for example, can be derived in FT via the Kubo formula
\begin{equation}
\eta = {\rm lim}_{\omega \rightarrow 0}\ \frac{1}{2\omega}\int dt\ d\vec{x}\ e^{i \omega t}\ \langle [T_{xy}(t,\vec{x}),T_{xy}(0,\vec{0})]\rangle\, .
\end{equation}
The gravity calculation for the $g_{xy}$ component of the metric determines the correlator and finally the value of the shear viscosity over entropy density \cite{pss}
\begin{equation}
\frac{\eta}{s}=\frac{1}{4\pi}\, .
\end{equation}
It is by now well known that this result for ${\cal N}=4$ SYM in the planar limit is surprisingly compatible with the experimental results at RHIC and LHC, even though experimental errors remain
large. Other methods of obtaining this quantity, such as perturbative QCD or lattice, give higher values, hardly compatible with experiments.

The result above shows both the relative simplicity of holographic methods and their possible relevance for a deeper understanding of experimental settings, whenever there are no reliable direct theoretical calculations available in QCD.
An extensive review of the applications of the gauge/gravity duality to the physics of
heavy ion collisions has appeared
in \cite{CasalderreySolana:2011us}.

\section{D3-D7 Quark-Gluon Plasmas}

In the heavy ion scattering experiments at RHIC and LHC a Quark-Gluon Plasma is produced. This is a thermalized system at finite (small) baryon chemical potential where the quarks and gluons are deconfined.
>From a theoretical point of view the investigation of the QGP is quite challenging.
In fact, the analysis of
the experiments indicates that it is in a strong coupling regime, rendering perturbative QCD not entirely suitable for the purpose.
Lattice methods are clearly one of the best instruments for this investigation, but they cannot be exhaustive.
Apart from the sign problem for chemical potential, which can be overcome in some way for small values,
the main difficulty of the lattice approach is that it is not very suitable for real time physics (transport coefficients, probe physics), which plays an essential role in the experimental systems.

In this context, the holographic approach can furnish novel insight in the problem.
It is automatically a strong coupling approach and, as we have seen above, it allows to access with ease real time phenomena.
Moreover, as we will review, it allows for the investigation of theories with dynamical flavors at finite chemical potential.
The obvious price we have to pay in this approach is that we are investigating a theory, in the planar limit and very large 't Hooft coupling, which is not QCD, but one of its relatives.

In this note, we are interested in reviewing the inclusion of dynamical flavors in the holographic approach to the physics of the ${\cal N}=4$ $SU(N_c)$ SYM plasma.
Flavors are studied by introducing $N_f$ explicit D7-branes in the $AdS$ black hole background.
In the strict 't Hooft limit ($N_c \rightarrow \infty$ with $N_f$ fixed), the
branes can be treated in the probe approximation, where their
backreaction on the background is ignored and therefore the flavors
are quenched \cite{kk}.
This approach allows to study a number of physical problems, but it misses part of the physics of the RHIC and LHC QGPs, where a significant fraction of degrees of freedoms are dynamical fundamental quarks.

The problem with going beyond the quenched approximation is that calculating the backreaction of the flavor branes is usually a very difficult task, involving the solution of systems of partial differential equations.
In order to overcome this difficulty, in \cite{noncritical,cnp} a method was introduced,\footnote{For a review of this approach, see \cite{npr}.} termed "smearing technique", which takes advantage of the parameter regime under consideration.
Since we are in the planar limit $N_c \gg 1$, the probe approximation breaks down when we introduce many flavor branes, $N_f \gg 1$.
Thus, these many branes can be homogeneously distributed in their transverse space, recovering (part of) the original symmetry of the solution and typically reducing the system to a set of ordinary differential equations.
The main effect in the dual field theory is that the considered flavor group is Abelian.

Employing this technique, the backreacted background corresponding to massless flavors can be derived in the form
\begin{equation}
ds_{10}^2 = h^{-1/2}[ -b\, dt^2 + dx^i dx_i]  + h^{1/2}[ bS^8 F^2  d\sigma^2 +S^2 ds^2_{KE} + F^2(d\tau + A_{KE})^2] \label{genansatz} \,  ,
\end{equation}
where "$KE$" indicates a K\"ahler-Einstein manifold, which in the ${\cal N}=4$ case is $CP_2$, and $A_{KE}$ is the connection one-form whose curvature is related to the K\"ahler form of the $KE$ base: $dA_{KE}= 2\,J_{KE}$. Using the invariance under diffeomorphisms we have chosen a radial coordinate $\sigma$ which is convenient to write the equations of motion of the system. The solution with zero charge has been derived in \cite{Bigazzi:2009bk}, while its generalization at finite charge appears in \cite{bis}.\footnote{The zero temperature solution for massless flavors appeared in \cite{Benini:2006hh}, the ones for massive flavors in \cite{fkw1,fkw2,Bigazzi:2009bk}.}
It can be expressed by means of two parameters
\begin{equation}
\epsilon_h \sim \lambda_h \frac{N_f}{N_c}\, , \qquad \qquad \hat \delta \sim \frac{1}{\sqrt{\lambda_h}} \frac{\mu}{T} \left(1 - \frac{5}{24}\epsilon_h \right)\, ,
\end{equation}
where $\mu$ is the quark chemical potential and the subindex $_h$
means that the coupling is evaluated at the temperature of the
plasma. Up to order $\epsilon_h^2, \hat \delta^2$ the perturbative
solution can be derived in analytic form and expressed in a
new\footnote{For the explicit form of the equations satisfied by the
fields in the ansatz in the variable $\sigma$, we refer to the
original paper \cite{bis}.} radial variable $r$ such that
$h=R^4/r^4$, $R$ being the radius of the undeformed $AdS$ space
\begin{eqnarray}
b(r) &=& 1-\frac{r_h^4}{r^4} - \frac{\epsilon_h\dent^2}{2} \left( \left(2-\frac{r_h^4}{r^4}\right)
\left(\frac{r_h^2}{r^2}-\log \left[1 + \frac{r_h^2}{r^2}\right]
\right) - \frac{r_h^4}{r^4} (1- \log 2)\right)
\nonumber\\
&& ~~~~~~~~~ + \frac{\epsilon_h^2\dent^2}{12} \left(
17\frac{r_h^2}{r^2} - \frac{29}{2}\frac{r_h^4}{r^4} - \frac{5}{2}
\frac{r_h^6}{r^6} - \frac{17}{2} \left(2 - \frac{r_h^4}{r^4} \right)\log\left[ 1+ \frac{r_h^2}{r^2}\right]
+ \frac{17}{2}  \frac{r_h^4}{r^4} \log2 \right)\,,\nonumber\\
S(r) &=&  r\left[ 1 +  \frac{\epsilon_h}{24} + \epsilon_h^2 \left( \frac{9}{1152} -
\frac{1}{24}\log \frac{r_h}{r}\right)   \right. \nonumber\\
&&  \quad+\frac{\epsilon_h\dent^2}{40}\left( 3 -2\frac{r^2}{r_h^2}
-3\left(1-2 \frac{r^4}{r_h^4} \right) \log\left[1 +
\frac{r_h^2}{r^2}\right]- \frac{1}{2} G(r)   \right)
\nonumber\\
&& \left. \quad +  \frac{\epsilon_h^2\dent^2}{320} \left(-33+ 22
\frac{r_h^2}{r^2} + 33  \left(1-2 \frac{r^4}{r_h^4}\right)
\log\left[1 + \frac{r_h^2}{r^2}\right]+ \frac{11}{2} G(r)
\right)\right]\,,\nonumber
 \\
F(r) &=&  r\left[  1 -  \frac{\epsilon_h}{24} + \epsilon_h^2 \left( \frac{17}{1152} +
\frac{1}{24}\log \frac{r_h}{r}\right)   \right. \nonumber\\
&& \quad +  \frac{\epsilon_h\dent^2}{40}\left( 3
-22\frac{r^2}{r_h^2} +5\frac{r_h^2}{r^2} -3 \left(1-2
\frac{r^4}{r_h^4} \right) \log\left[1 + \frac{r_h^2}{r^2}\right] +
2G(r)   \right)
\nonumber\\
&& \left. \quad +  \frac{\epsilon_h^2\dent^2}{192} \left(-21+ 154\frac{r^2}{r_h^2}
-35 \frac{r_h^2}{r^2} + 21 \left(1-2\frac{r^4}{r_h^4}\right) \log\left[1+ \frac{r_h^2}{r^2}\right]
- 14\, G(r)  \right) \right]\, , \nonumber\\
\Phi(r) &=& \Phi_h + \epsilon_h\log\frac{r}{r_h}  -
\frac{\epsilon_h^2}{48}\left(8 \left(1+3\log \frac{r}{r_h}\right) \log\frac{r_h}{r}
- 3 \,Li_2\left[1-\frac{r_h^4}{r^4}\right]\right) \nonumber\\
     &&\quad + \frac{\epsilon_h^2\dent^2}{120}\left( 41 - 2\pi - 26 \frac{r^2}{r_h^2}  - 15\frac{r_h^2}{r^2} +
     G(r)+  \left(11 + 18\frac{r^4}{r_h^4}\right) \log\left[1 + \frac{r_h^2}{r^2}\right]-29\log 2  \right)\, ,
   \nonumber
\end{eqnarray}
 where $\Phi$ is the dilaton, $G(r)=2\pi
\frac{r_h^6}{r^6}\,{}_2F_{1}\left(\frac{3}{2},\frac{3}{2},1,1-\frac{r_h^4}{r^4}\right)$
is an hypergeometric function and $Li_2(u)\equiv \sum_{n=1}^\infty
\frac{u^n}{n^2}$ is a polylogarithmic function. The parameter
$r_{h}$ marks the (perturbative) position of the horizon defined by
$b(r_h) = 0 +{\cal O}(\epsilon_h^3,\hat \delta^4)$. In addition to the
metric and the dilaton there are non-trivial differential forms and
a $U(1)$ field turned on in the worldvolume of the D7-branes, dual
to the chemical potential. We refer to \cite{bis} for details.

The regime of validity of this solution is limited to the usual range $N_c\gg 1, \lambda_h\gg 1$ and,
as we have seen, $N_f\gg 1$; moreover, since the theory has a Landau pole in the UV, in order to decouple
the IR physics from the UV we must require $\epsilon_h \ll 1$; finally, in order to have an analytic
solution we required $\dent \ll 1$, an assumption that in principle can be relaxed.

\section{Results}

The solution described in the previous section allows us to study a number of effects of dynamical flavors at finite baryon charge in a strongly coupled theory in a completely controllable setting.
As a first topic, let us exhibit the thermodynamic quantities (entropy density $s$, energy density $\varepsilon$, grand potential $\omega$, specific heat $c_v$) in the grand canonical ensemble
\begin{eqnarray}
s&=&\frac12 \pi^2 N_c^2 T^3 \left[1+\frac12 \epsilon_h (1+  \dent^2)
+\frac{7}{24}\epsilon_h^2 (1+  \dent^2) \right]\, ,\\
\varepsilon &=& \frac38 \pi^2 N_c^2 T^4
\left[1+\frac12 \epsilon_h(1+ 2 \dent^2) + \frac13 \epsilon_h^2(1+ \frac74 \dent^2)\right]\, ,\\
\omega &=&-p=-\frac18 \pi^2
 N_c^2 T^4 \left[1+\frac12 \epsilon_h (1+ 2 \dent^2)+\frac16 \epsilon_h^2(1+ \frac72 \dent^2)\right]\, ,\\
c_v &=& \frac{3}{2} \pi^2  N_c^2 T^3   \left[ 1+ \frac{1}{2} \epsilon_h \left( 1+\dent^2 \right) + \frac{1}{24}\epsilon_h^2 \left( 11+7\dent^2 \right) \right]\, .     \nonumber
\end{eqnarray}
Note that in order for the standard thermodynamic relations to hold it is essential that the following dependence is actually realized: $\frac{d\epsilon_h}{dT}=\frac{\epsilon_h^2}{T}+O(\epsilon_h^3)$, $\left( \frac{d \, \dent}{d T}\right)_\mu = - \frac{\dent}{T} \left( 1 +\frac{\epsilon_h}{2}  + O(\epsilon_h^2) \right)$.

The breaking of conformality due to fundamental matter is a second order effect in $\epsilon_h$ and is not affected by the finite charge density
\begin{eqnarray}
\varepsilon-3p&=&\frac{1}{16}\pi^2 N_c^2 T^4 \epsilon_h^2\, ,\\
c_s^2 &=&  \frac13 \left[1-\frac{1}{6} \epsilon_h^2\right]\, ,
\end{eqnarray}
where $c_s$ is the speed of sound.

Another example of the interest of such kind of solution is provided
by the study of energy loss of a probe parton in the plasma. It is
known  that in the experimental setting the energy loss is huge and
it is very interesting to characterize such process from a
theoretical point of view. We will concentrate on one coefficient,
the "jet quenching parameter" $\hat q$ which accounts for such a
characterization. It is defined as the transverse momentum squared
transferred by the plasma to the parton per mean free path. In
string theory there is a very simple way of calculating such
quantity by means of a partially light-like Wilson loop \cite{lrw}.
In the case at hand the outcome of this calculation is (neglecting
$O(\epsilon_h^2)$ terms) \cite{Bigazzi:2009bk,bis}
\begin{equation}\label{jet}
\hat q=\frac{\pi^{3/2}\Gamma(\frac34)}{\Gamma(\frac54)}\sqrt{\lambda_h}\,T^3
\left[1 + \frac{2+\pi}{8} \epsilon_h (1+0.8675\, \dent^2)
\right]\, .
\end{equation}
The naive interpretation of this formula is that both uncharged
flavors and a finite charge enhance the jet quenching, i.e. the
energy loss of the parton.\footnote{In the uncharged case, this
result was already obtained in \cite{noncrittermo} from non-critical
string models, which on the other hand suffer from uncontrolled
approximations and so are not quantitatively fully reliable.}
Although being a priori completely unjustified, one can plug in the
formula above quantities reasonable for the RHIC experiment, such as
$N_c=N_f=3$, $T=300$ MeV, $\lambda_h = 6\pi$, $\mu=10$ MeV,
obtaining $\hat q \sim 5.3\ (GeV)^2/fm$ \cite{Bigazzi:2009bk}, as
opposed to $\hat q \sim 4.5\ (GeV)^2/fm$ for the unflavored theory
\cite{lrw}. The most common values reported for RHIC are $\hat q
\sim 5-15\ (GeV)^2/fm$, so the naive extrapolation of the formula
(\ref{jet}) is surprisingly close to the real values. This fact
could be a total accident, or else a signal that the energy loss
process at strong coupling is quite insensitive to the details of
the theory under consideration, as it happens for the shear
viscosity over entropy density value.

The above statement of the enhancement of the jet quenching due to flavors and charge density is definitively too naive, since it involves the comparison of two theories with different numbers of degrees of freedom.
Adding fundamental flavors and charge enhances the entropy density, so the enhancement of the parton energy loss could be a trivial consequence of this fact.
In order to disentangle the two effects, one can compare the unflavored and flavored theories keeping fixed the number of degrees of freedom, i.e. either the entropy or the energy density.
This can be achieved by adjusting the temperature $T$, obtaining the result that flavors do enhance the jet quenching but the charge density reduces this enhancement.
Or, one can fix the number of degrees of freedom by varying $N_c$ (at fixed $T$).
In the latter case, while again the flavors enhance the jet quenching, the effect of the finite charge is to increase the enhancement.

All in all, the solid conclusion that can be
extracted from the result above is that dynamical flavors enhance the jet
quenching; note that this is probably the only reliable computation of the
effects of dynamical flavors at strong coupling on the energy loss of
partons in a plasma.
Considering that flavor effects are often discarded as subleading in phenomenological estimates \cite{muller}, the result should give an important indication that the latter approximation is probably not justified.

\section{On hydrodynamics of holographic plasmas}

There are solid lattice indications that the QCD plasma is both nearly conformal and strongly coupled in the temperature window relevant for the present experiments $1.5 T_c \leq T \leq 4T_c$.
Probably the simplest way of modeling this situation holographically is by an $AdS$ background deformed by a scalar dual to a marginally relevant operator.
The latter theory behaves effectively, at leading order in the deformation, as a so-called Chamblin-Reall model.
For these theories, all the hydrodynamic transport coefficients up to second order can be extracted from the results in \cite{Kanitscheider:2009as}.
Thus, one can give an estimate of the (initial behavior of) transport coefficients, up to second order, for RHIC and LHC \cite{noi,stima}.
With the definition\footnote{Note that $\delta$ and $\dent$ are completely unrelated terms.}
\begin{equation}\label{Delta}
\delta \equiv 1-3c_s^2\,,
\end{equation}
where $c_s$ is the speed of sound, and referring to the hydrodynamic stress-energy tensor in (\ref{enmom}),
the transport coefficients are given in Table \ref{relations}.\footnote{The uncharged flavored ${\cal N}=4$ SYM plasma has these same coefficients with $\delta=\epsilon_h^2/6$. Unfortunately, the Chamblin-Reall model is not able to describe the charged case due to the presence of extra fields in the gravitational setup.}
Considering the difficulty of dealing with such coefficients in QCD, this information could be useful in numerical simulations of the hydrodynamic evolution of the QGP.
In particular, the behavior with the temperature and the speed of sound of the shear and bulk relaxation times $\tau_\pi, \tau_\Pi$ is both potentially relevant and unexpected.
\begin{table}[h]\label{relations}
\begin{center}
\begin{tabular}{||c|c||c|c||c|c||}
\hline
 & & & & & \\
$ \frac{\eta}{s} $ & $\frac{1}{4\pi}$ &  $T\tau_{\pi}  $  & $ \frac{2-\log{2}}{2\pi} + \frac{3(16-\pi^2)}{64\pi}\delta $  & $ \frac{T\kappa}{s} $  &  $  \frac{1}{4\pi^2}\Bigl(1-\frac34 \delta \Bigr) $  \\
 & & & & & \\
\hline \hline
 & & & & & \\
$\frac{T \lambda_1}{s}  $ & $\frac{1}{8\pi^2}\Bigl(1+\frac34 \delta \Bigr) $ & $\frac{T \lambda_2}{s} $ & $-\frac{1}{4\pi^2}\Bigl( \log{2}+\frac{3\pi^2}{32}\delta \Bigr) $ & $\frac{T \lambda_3}{s} $ & $0 $ \\
 & & & & & \\
\hline \hline
 & & & & & \\
$\frac{T\kappa^*}{s} $ & $-\frac{3}{8\pi^2}\delta $ & $T\tau_{\pi}^* $ & $-\frac{2-\log{2}}{2\pi}\delta $ & $\frac{T \lambda_4}{s}  $ & $0 $ \\
 & & & & & \\
\hline \hline
 & & & & & \\
$\frac{\zeta}{\eta} $ & $\frac23 \delta $ & $T\tau_{\Pi} $ & $\frac{2-\log{2}}{2\pi} $ & $\frac{T \xi_{1}}{s} $ & $\frac{1}{24\pi^2}\delta $ \\
 & & & & & \\
\hline \hline
 & & & & & \\
$ \frac{T \xi_{2}}{s} $ & $\frac{2-\log{2}}{36\pi^2}\delta $ & $\frac{T \xi_{3}}{s} $ & $0 $ & $\frac{T \xi_{4}}{s} $ & $0 $ \\
 & & & & & \\
\hline \hline
 & & & & & \\
$\frac{T \xi_{5}}{s} $ & $\frac{1}{12\pi^2}\delta $ & $\frac{T \xi_{6}}{s} $ & $\frac{1}{4\pi^2}\delta $ & & \\
 & & & & & \\
\hline
\end{tabular}
\end{center}
\caption{The transport coefficients at leading order in the conformality deformation parameter $\delta \equiv 1-3c_s^2$.}\label{relations}
\end{table}

\section{Future directions}
The technique employed above in order to study the dynamics of fundamental flavors is suitable for a number of possible interesting applications in the near future.
Clearly, a first route would be to extend the analysis above beyond the small $\dent$ regime, exploring the large chemical potential region of the phase diagram which could correspond to extremality of the dual black hole.
Moreover, considering the results above, a deeper investigation of the probe parton physics would be quite important, considering also the experimental possibilities of the LHC experiment.
It would be also interesting to explore the optical properties of the system along the lines of \cite{ragazzi1,ragazzi2}.
Finally, it would be worth exploring the condensed matter applications of this formalism in the context of the holographic duality.

\acknowledgments We thank Javier Mas for his suggestions on this
note. The research leading to the results in this paper has has
received funding from the European Community Seventh Framework
Programme (FP7/2007-2013 under grant agreements n. 253534 and
253937), the Netherlands Organization for Scientific Research (NWO)
under the FOM Foundation research program, the grants
FPA2007-66665C02-02 and DURSI 2009 SGR 168, and the CPAN
CSD2007-00042 project of the Consolider-Ingenio 2010 program.


\begin{thebibliography}{99}

\bibitem{malda}
  J.~M.~Maldacena,
  \emph{The large N limit of superconformal field theories and supergravity,
  Adv.\ Theor.\ Math.\ Phys.\ } {\bf 2}, 231 (1998)
  [\emph{Int.\ J.\ Theor.\ Phys.\ } {\bf 38}, 1113 (1999)]
  [arXiv:hep-th/9711200].

\bibitem{gkp}
  S.~S.~Gubser, I.~R.~Klebanov and A.~M.~Polyakov,
  \emph{Gauge theory correlators from non-critical string theory,
  Phys.\ Lett.\  B }{\bf 428}, 105 (1998)
  [arXiv:hep-th/9802109].

\bibitem{witten}
  E.~Witten,
  \emph{Anti-de Sitter space and holography,
  Adv.\ Theor.\ Math.\ Phys.\ } {\bf 2}, 253 (1998)
  [arXiv:hep-th/9802150].

\bibitem{baier}
  R.~Baier, P.~Romatschke, D.~T.~Son, A.~O.~Starinets and M.~A.~Stephanov,
  \emph{Relativistic viscous hydrodynamics, conformal invariance, and holography,
  JHEP }{\bf 0804}, 100 (2008)
  [arXiv:0712.2451 [hep-th]].
S.~Bhattacharyya, V.~E.~Hubeny, S.~Minwalla and M.~Rangamani,
  \emph{Nonlinear Fluid Dynamics from Gravity,
  JHEP }{\bf 0802}, 045 (2008)
  [arXiv:0712.2456 [hep-th]].
 M.~Natsuume and T.~Okamura,
  \emph{Causal hydrodynamics of gauge theory plasmas from AdS/CFT duality,
  Phys.\ Rev.\  D }{\bf 77}, 066014 (2008)
  [Erratum-ibid.\  D {\bf 78}, 089902 (2008)]
  [arXiv:0712.2916 [hep-th]].

\bibitem{romatschke}
  P.~Romatschke,
  \emph{Relativistic Viscous Fluid Dynamics and Non-Equilibrium Entropy,
Class.\ Quant.\ Grav.\  }{\bf 27}, 025006 (2010)
  [arXiv:0906.4787 [hep-th]].

\bibitem{pss}G.~Policastro, D.~T.~Son and A.~O.~Starinets,
  \emph{The shear viscosity of strongly coupled N = 4 supersymmetric Yang-Mills
  plasma,
  Phys.\ Rev.\ Lett.\  }{\bf 87}, 081601 (2001)
  [arXiv:hep-th/0104066].

\bibitem{CasalderreySolana:2011us}
  J.~Casalderrey-Solana, H.~Liu, D.~Mateos, K.~Rajagopal and U.~A.~Wiedemann,
  \emph{Gauge/String Duality, Hot QCD and Heavy Ion Collisions,}
  arXiv:1101.0618 [hep-th].

\bibitem{kk}
  A.~Karch and E.~Katz,
  \emph{Adding flavor to AdS/CFT,
  JHEP }{\bf 0206}, 043 (2002)
  [arXiv:hep-th/0205236].

\bibitem{noncritical}
  F.~Bigazzi, R.~Casero, A.~L.~Cotrone, E. Kiritsis and A. Paredes,
  \emph{Non-critical holography and four-dimensional CFT's with fundamentals,
  JHEP }{\bf 0510}, 012 (2005)
  [hep-th/0505140].

\bibitem{cnp}
  R.~Casero, C.~Nunez and A.~Paredes,
  \emph{Towards the string dual of N = 1 SQCD-like theories,
  Phys.\ Rev.\  D }{\bf 73}, 086005 (2006)
  [arXiv:hep-th/0602027].

\bibitem{npr}
  C.~Nunez, A.~Paredes and A.~V.~Ramallo,
  \emph{Unquenched flavor in the gauge/gravity correspondence,
  Adv.\ High Energy Phys.\  }{\bf 2010}, 196714 (2010)
  [arXiv:1002.1088 [hep-th]].

\bibitem{Bigazzi:2009bk}
  F.~Bigazzi, A.~L.~Cotrone, J.~Mas, A.~Paredes, A.~V.~Ramallo and J.~Tarrio,
  \emph{D3-D7 Quark-Gluon Plasmas,
  JHEP }{\bf 0911}, 117 (2009)
  [arXiv:0909.2865 [hep-th]].

\bibitem{bis}
  F.~Bigazzi, A.~L.~Cotrone, J.~Mas, D. Mayerson and J.~Tarrio,
  \emph{D3-D7 Quark-Gluon Plasmas at Finite Baryon Density,}
   arXiv:1101.3560 [hep-th].

\bibitem{Benini:2006hh}
 F.~Benini, F.~Canoura, S.~Cremonesi, C.~Nunez and A.~V.~Ramallo,
 \emph{Unquenched flavors in the Klebanov-Witten model,
 JHEP }{\bf 0702}, 090 (2007)
 [arXiv:hep-th/0612118].

\bibitem{fkw1} F.~Bigazzi, A.~L.~Cotrone and A.~Paredes,
 \emph{Klebanov-Witten theory with massive dynamical flavors,
 JHEP }{\bf 0809}, 048 (2008)
 [arXiv:0807.0298 [hep-th]].

\bibitem{fkw2}F.~Bigazzi, A.~L.~Cotrone, A.~Paredes and A.~V.~Ramallo,
 \emph{Non chiral dynamical flavors and screening on the conifold,
Fortsch.\ Phys.\ } {\bf 57}, 514 (2009)
 [arXiv:0810.5220 [hep-th]].

\bibitem{lrw}H. Liu, K. Rajagopal and U. A. Wiedemann,
\emph{Calculating the jet quenching parameter from AdS/CFT,
Phys.\ Rev.\ Lett.\ } {\bf 97}, 182301 (2006)
[arXiv:hep-ph/0605178].
F.~D'Eramo, H.~Liu and K.~Rajagopal,
  \emph{Transverse Momentum Broadening and the Jet Quenching Parameter, Redux,}
  arXiv:1006.1367 [hep-ph].

\bibitem{noncrittermo}G.~Bertoldi, F.~Bigazzi, A.~L.~Cotrone and J.~D.~Edelstein,
 \emph{Holography and Unquenched Quark-Gluon Plasmas,
 Phys.\ Rev.\  D }{\bf 76}, 065007 (2007)
 [arXiv:hep-th/0702225].

\bibitem{muller}
 B.~Muller and J.~L.~Nagle,
 \emph{Results from the Relativistic Heavy Ion Collider,
 Ann.\ Rev.\ Nucl.\ Part.\ Sci.\ } {\bf 56}, 93 (2006)
 [arXiv:nucl-th/0602029].

\bibitem{Kanitscheider:2009as}
  I.~Kanitscheider and K.~Skenderis,
  \emph{Universal hydrodynamics of non-conformal branes,
  JHEP }{\bf 0904}, 062 (2009)
  [arXiv:0901.1487 [hep-th]].

\bibitem{noi}F.~Bigazzi, A.~L.~Cotrone and J.~Tarrio,
  \emph{Hydrodynamics of fundamental matter,
 JHEP }{\bf 1002}, 083 (2010)
  [arXiv:0912.3256 [hep-th]].

\bibitem{stima}F.~Bigazzi and A.~L.~Cotrone,
  \emph{An elementary stringy estimate of transport coefficients of large temperature QCD,
  JHEP} {\bf 1008}, 128 (2010).
  [arXiv:1006.4634 [hep-ph]].

\bibitem{ragazzi1}
  A.~Amariti, D.~Forcella, A.~Mariotti and G.~Policastro,
  \emph{Holographic Optics and Negative Refractive Index},
  arXiv:1006.5714 [hep-th].

\bibitem{ragazzi2}
A.~Amariti, D.~Forcella and A.~Mariotti,
  \emph{Additional Light Waves in Hydrodynamics and Holography},
  arXiv:1010.1297 [hep-th].


\end{thebibliography}
\end{document}